\documentclass[aps,prb,preprint,superscriptaddress]{revtex4-1}
\usepackage{newfloat}
\DeclareFloatingEnvironment[name={Fig. S}]{suppfigure}

\usepackage{graphicx}
\usepackage{amsmath}
\usepackage{amsfonts}

\usepackage{graphicx}
\usepackage{dcolumn}
\usepackage{bm}
\usepackage{amssymb}

\begin{document}

\title{Spin-orbital polarons in electron doped copper oxides}

\author{Anna Kusmartseva}
\affiliation{Department of Physics, Loughborough University,
LE11 3TU Loughborough, United Kingdom}

\author{Heshan~Yu}
\affiliation{Beijing National Laboratory for Condensed Matter Physics, Institute of Physics, Chinese Academy of Sciences, Beijing 100190, China}

\author{Kui Jin}
\affiliation{Beijing National Laboratory for Condensed Matter Physics, Institute of Physics, Chinese Academy of Sciences, Beijing 100190, China}
\affiliation{\mbox{Collaborative Innovation Center of Quantum Matter, Beijing, 100190, China}}

\author{F. V. Kusmartsev}
\affiliation{Department of Physics, Loughborough University,
LE11 3TU Loughborough, United Kingdom}

\begin{abstract}
Present work demonstrates the formation of spin-orbital polarons in electron doped copper oxides, that arise due to doping-induced polarisation of the oxygen orbitals in the CuO$_2$ planes. The concept of such polarons is fundamentally different from previous interpretations. The novel aspect of spin-orbit polarons is best described by electrons becoming self-trapped in one-dimensional channels created by polarisation of the oxygen orbitals. The one-dimensional channels form elongated filaments with two possible orientations, along the diagonals of the elementary CuO$_2$ square plaquette. As the density of doped electrons increases multiple filaments are formed. These may condense into a single percollating filamentary phase. Alternatively, the filaments may cross perpendicularly to create an interconnected conducting quasi-one-dimensional web. At low electron doping the antiferromagnetic (AFM) state and the polaron web coexist. As the doping is increased the web of filaments modifies and transforms the AFM correlations leading to a series of quantum phase transitions - which affect the normal and superconducting state properties.  

 \end{abstract}
\maketitle
\pagenumbering{arabic}
\pagestyle{plain}







Multiple experimental evidences show that electron-doped cuprates are substantially different from their hole-doped counterparts\cite{Ishii-2014,Jin-2009,MIT-1998,Dagan-2004,Heshan-2017,Jin-2015}. 
Both systems are characterized by strongly correlated electrons and can be described by Hubbard-like models. The main differences arise with doping. Here, the main focus is given to lanthanum cuprates where electron and hole doping can be achieved by substituting Ce or Sr on the La sites of the parent compound La$_2$CuO$_4$  respectively. Electronically, La$_2$CuO$_4$ corresponds to an AFM state associated with a half-filled Hubbard band\cite{Zhang-1988,hlubina1999phase}. When the system is doped with holes, the resulting dominant electronic correlations are confined to the lower Hubbard band. On the other hand with electron doping the upper or second Hubbard band becomes more relevant. Furthermore, a mechanism is suggested to show how the addition of electrons may induce polarisation of the oxygen orbitals in the CuO$_2$ plane.

\begin{figure}
\begin{center}
\includegraphics[width=15cm,height=9cm]{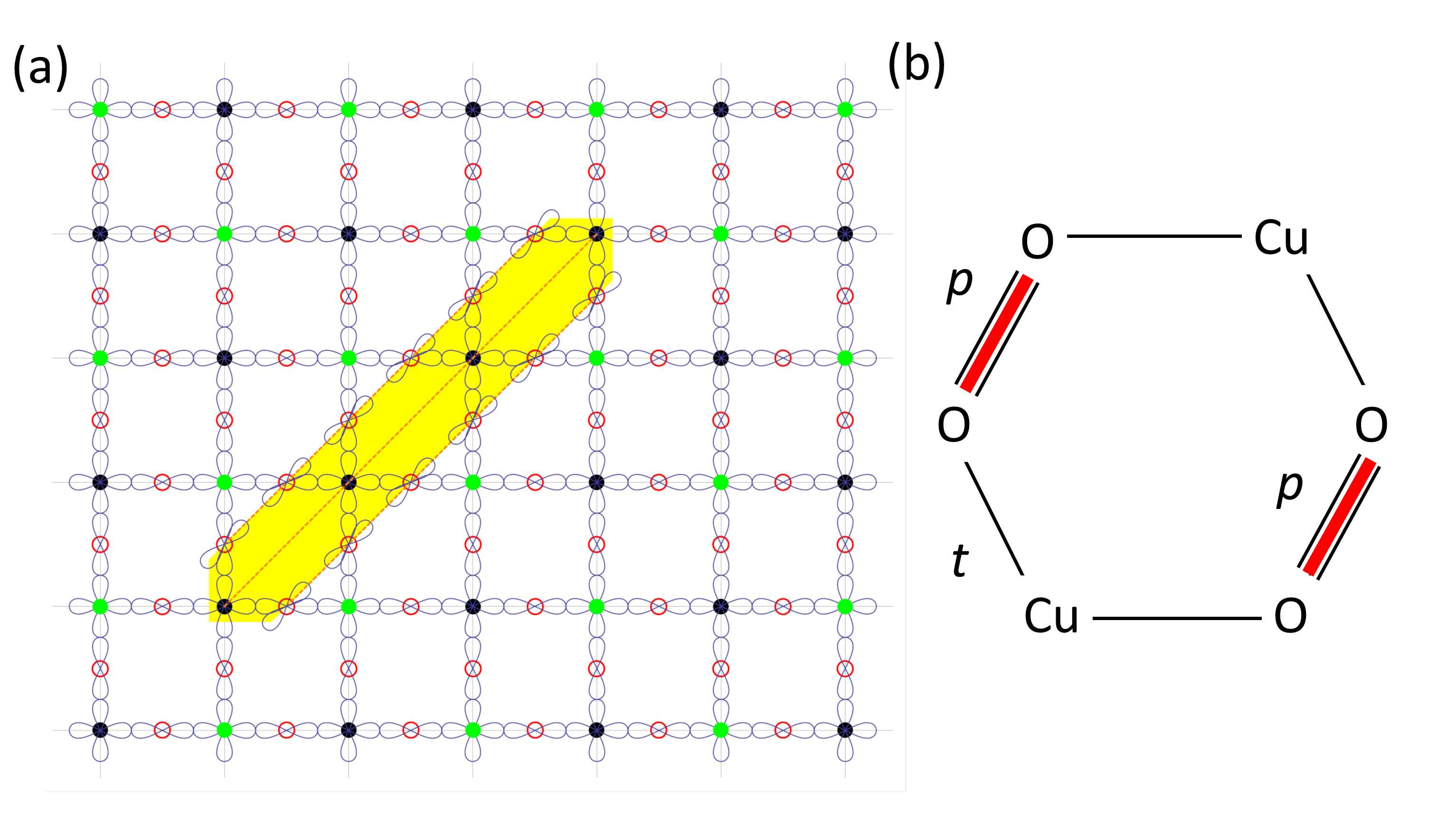}
\caption{\label{spin-polaron} a) A schematic representation of the spin-orbital polaron created in the CuO$_2$ plane is shown. Its position is associated with the yellow polygon. The polaron arises due to the polarisation of the $p_x$ and $p_y$ orbitals on the oxygen atoms neighbouring the Cu sites by the doped electron. These polarised orbitals are denoted by the red dashed line.  Here, for example, the polaron is localized across 4 distinct spin-up Cu sites (black points inside the yellow area). Usually, the polaron has elongated, quasi one-dimensional character and a needle-like shape. Within the polaron area the AFM order parameter  $\Delta_{SDW}$  decreases. b) The lowest order and smallest size polaron occupies a hexagonal plaquette consisting of 4 oxygen and 2 copper atoms. A tight-binding model, involving the hybridisation and hopping integrals $p$ and $t$ respectively, is successfully used to describe the electronic correlations across these 6 sites.}
\end{center}
\end{figure}	

In the parent compound La$_2$CuO$_4$ due to band half-filling the AFM order is attributed to half of the Cu sites having spin-up and the other half having spin-down orientations in the CuO$_2$ plane. The main difference is that with hole doping some 
copper sites occupied by holes act effectively as Jahn-Teller ions \cite{Muller-1987,Kusmartsev-2015}, which creates an interesting possibility for the formation of anti-ferroelectricity in hole-doped cuprates\cite{Kusmartsev-2015,Saarela-2017}. On the other hand, electron doping may result in completely filled Cu orbitals with electronic $d^{10}$ state corresponding to the upper Hubbard band. In contrast to the hole-doped case, the formation of the  $d^{10}$ configuration is not associated with a Jahn-Teller lattice distortion.  

Another key difference is that due to exchange forces the doped electron can only occupy copper sites with opposite spin orientation, which costs the large Hubbard energy $U\sim 10$ eV. 
Here the situation with electron doping is similar to the Zhang-Rice singlet in hole-doped cuprates 
\cite{Zhang-1988}.  Specifically, due to  the
large on-site Hubbard repulsion $U$ 
the charge density cloud of the doped electron will shift away from the original Cu site, spreading out onto the neighbouring oxygen sites. This process will strongly polarise the oxygen orbitals, particularly those not directly involved in $p-d$ bonding. The orbital polarisation may extend to next-neighbouring oxygens resulting in the formation of a polarisation potential well spanning over several Cu sites with the same spin orientation (Fig.\ref{spin-polaron}a). The doped electron may become self-trapped by this polarisation well. Notably, the spin of the doped electron is different to the one located on the Cu sites inside the polarisation cloud.

To describe the formation of the spin-orbital polaron a schematic of the CuO$_2$ plane is considered, where the Cu spin-up and spin-down sites are marked by black and green points respectively, while the oxygen atoms are represented by open red circles (see, Fig.\ref{spin-polaron}a). The electron cloud of the self-trapped doped electron extends from a single Cu site to the neighbouring oxygen atoms. The $p_x$ or $p_y$ oxygen orbitals which are perpendicular to the $p-d$ Cu-O bonding become also polarised. Note, the polarisation is oriented along the diagonal in the CuO$_2$ plane (Fig. \ref{spin-polaron}a - red dashed line). In addition, the $p_x$ and $p_y$ orbitals may hybridise weakly to contribute to a non-zero, substantial overlapping integral, $p$ (Fig.\ref{spin-polaron}b). The combined orbital hybridisation and polarisation form a potential well for the doped electron, where it may become self-trapped over several Cu sites with the same spin orientation. Figure \ref{spin-polaron}a shows a polaron (yellow coloured area) trapped over four spin-up Cu sites. The size of the smallest polaron involves two copper and four oxygen atoms spanning over a hexagonal plaquette (Fig.\ref{spin-polaron}b). It is important to note that problems related to orbital and spin-orbital polarons formation as well as the creation of polaron strings are rather topical and were actively discussed in the past, see e.g. refs\cite{Mizokawa-2001,Wohlfeld-2009a,Wohlfeld-2009}. For example, the formation of spin-polarons in La$_2$CuO$_4$ was shown to originate with hole doping into the AFM ground state \cite{4,5}. In transition metal oxides, specifically LaMnO$_3$, orbital polarons were equally connected to hole doping of the orbital ground state. Here, all these ideas are combined to consider the creation of spin-orbital polarons arising due to oxygen orbital polarisation induced by electron doping, which is conceptually different from the case discussed in LaMnO$_3$ and other systems\cite{Mizokawa-2001,Wohlfeld-2009a,Wohlfeld-2009,4,5,Saarela-2015}.

The formal description of the mechanism for the spin-orbital polaron and the conducting filament formation begins with the Heisenberg AFM state of the parent compound La$_2$CuO$_4$ which is well characterized by the Hubbard model at half-filling. To take into the account the polarisation of the oxygen orbitals the Zhang-Rice model\cite{Zhang-1988} is introduced: 

\begin{equation}
H =  \sum_{i, \sigma}  \epsilon_d C_{i\sigma}^\dagger C_{i\sigma} + \sum_{i, \sigma}  \epsilon_p p_{i\sigma}^\dagger p_{i\sigma} + t \sum_{<ij>} C_{i\sigma}^\dagger p_{j\sigma} +hc+ U  \sum_{i} n_{i\uparrow}  n_{i\downarrow}
\label{Hubbard-Ham}
\end{equation}
Here, the creation and annihilation operators of the doped electrons on the Cu and oxygen sites are denoted by $C_{i\sigma}^\dagger$, $C_{j\sigma}$, $p_{i\sigma}^\dagger$ and $ p_{j\sigma}$  respectively.  $t$ is the hopping integral between the Cu and O sites. The occupation number operators for spin up and spin down electrons are given by  $n_{i\uparrow}=C_{i\uparrow}^\dagger C_{j\uparrow}$ and  $n_{i\downarrow}= C_{i\downarrow}^\dagger C_{j\downarrow}$, respectively, while $U$ is the on-site Hubbard repulsion constant.

When describing the case of hole-doped cuprates Eq. \ref{Hubbard-Ham} naturally transforms into the $t-J$ model, where the double occupancy of the copper sites is prohibited and the Zhang-Rice singlet is formed \cite{Zhang-1988}. The situation is different when considering electron-doped cuprates. The most obvious distinction is that the second Hubbard band becomes now filled, which by definition implies doubly occupied Cu states at the cost of the Hubbard $U$. However, a doped electron can partially avoid the double occupancy by displacing its electron density to the neighbouring oxygens, resulting in two different effects.
Firstly, this doping may induce a structural transition from the $T$ to the $T'$ phase which has no apex oxygens. The second effect is related to the polarisation of free (non-bonded) oxygen orbitals triggered by the displacement of the charge density from the Cu site.
Consequently if the charge deformation or the polarisation of the oxygen orbitals is taken into account, one may show that the hopping integral between the neighbouring oxygen sites increases.
It is naturally assumed that the polarised orbitals on the neighbouring oxygens in the CuO$_2$ plane lie along diagonals and become weakly hybridised. This leads to the appearance of a hopping path of the form $Cu-O-O-Cu$ where two next-neighbouring Cu sites are connected via two oxygen atoms (Fig. \ref{spin-polaron}b). Such process provides substantial probability for the tunnelling of the doped electron between the next neighbouring Cu sites. In this case the original Hubbard model should be complemented by the term:
\begin{equation}
H_2 =  \sum_{<<ij>>}  t'_{ij} p_{i\sigma}^\dagger p_{j\sigma} +hc 
\label{Hubbard2-Ham}
\end{equation}
where the value $t'_{ij}$ is the hopping integral between the nearest oxygen sites arising only in the local presence of the doped electron and within the area where its electron density is confined. In general, if there is any hopping between the oxygen atoms in the plane its value is usually considered to be significantly smaller than $t$, that is $t'_{ij}=t'\sim 0.1 t$. 
The proposed model is conventionally named as $t-t'$ Hubbard model and its phase diagram has been intensively discussed \cite{hlubina1999phase}. Here, the focus is shifted to the possible local enhancement of the $t'$ term arising due to the local polarisation on the oxygen. In this case the hopping integral on the bonds, $<<ij>$, where the oxygen orbital polarisation occurs (e.g. $p_x$ and $p_y$ ), becomes  equal to the value $t'_{ij}=p$. Hence, the hopping term is substantially enhanced from its original, non-polarised value $p>>t'$.
In particular, it is proposed that the wave packet or charge density of the doped electron involved in polarising the oxygen orbitals definitely contributes to the value of $t'$ hopping. The main effect is related to the formation of virtual hopping chains of the form $Cu-O-O-Cu$ which connect two Cu sites neighbouring along the diagonal of a CuO hexagonal plaquette. The polarised oxygen orbitals are key in this mechanism. 

Importantly, the polarisation energy of the system is equal to $E_p=\alpha p^2$.  Here $\alpha$ is a constant describing the polarisation rigidity of the system. It is playing the same role as elastic modulus in deformation energy of solids.  Below we follow the logistics of the Pekar polaron theory (see, recent extensive review and the references therein\cite{dykman2015roots}). We add the polarisation energy $E_p$ to the lowest eigenvalue $E(p)$ obtained from the solution of the Shr\"odinger equation $(H+H_2(p))\Psi=E(p)\Psi$ .  Therewith we obtain
the dependence of the polaron energy, $ J_{Pekar}(p)=E(p)+E_p$, on the parameter $p$, describing the polarisation of the oxygen orbitals.
In the simplest case of the smallest polaron the wave function, $\Psi$, describes an electron located on two copper and four oxygen sites, forming a hexagonal plaquette, see, the Fig.1b, superposed on an AFM background.
The solutions of this equation gives the polaron energy, $ J_{Pekar}(p)=E(p)+E_p$, which takes the form:
\begin{equation}
J_{Pekar}(p)=\frac{1}{2} \left(-\sqrt{p^2+8 t^2}-p\right)+ \alpha p^2
\label{Pekar}
\end{equation}
here, for simplicity it is assumed that $\epsilon_p=\epsilon_d$ and $U=0$. The largest eigenvalue is most significantly affected by changes in these parameters. 
The function $J_{Pekar}(p)$ has a clear minimum, the position of which $p_{min}$ depends on the values of $t$ and $\alpha$. For example, when $\alpha =0.256/t$ the value $p_{min}\approx 1.41t$. Although, this behaviour has an exact analytic expression, it is too expansive and non-compact to be included here.
 Instead, the approximation for $p_{min}$ can be written as: 
 \begin{equation}
 p_{min} \approx 1/4\alpha+\sqrt{(1/(1+32 t^2 \alpha^2))}/4\alpha,
 \label{pmin}
\end{equation}
which becomes exact at large values of $32 t^2 \alpha^2>1$. From Eq.\ref{Pekar} it becomes obvious that the zero polarisation state, $p=0$, is unstable and its energy decreases when the parameter $p$ increases from zero to $p_{min}$. Note, this is a universal phenomenon that is related to the polarisation of the oxygen orbitals. 
The value of the Hubbard $U$ associated with the on-site Coulomb repulsion at Cu sites has little to no effect on the position of the polarisation minimum $p_{min}$. This is  because the electron density is mostly located around the four oxygen sites with little remaining around the Cu positions (Fig. \ref{spin-polaron}a).  In the ground state associated with the lowest eigenvalue the electron density shifts from the Cu sites towards oxygen sites when the value of the parameter $p$ increases. When $p>1$ the polaron charge density is mostly located on the oxygen sites. Therefore the on-site Coulomb Hubbard repulsion at Cu sites $U$ has almost no influence on its energy.  Also the spin-orbital polaron is analogous to  other types of spin polarons\cite{Kriv,Naga,Schrieffer-1988,Mott1970-RPP}. It has an opposite spin to the spins of the Cu sites and its orientation is elongated along the diagonal of the CuO$_2$ square. 

\begin{figure}
\begin{center}
\includegraphics[width=15cm,height=9cm]{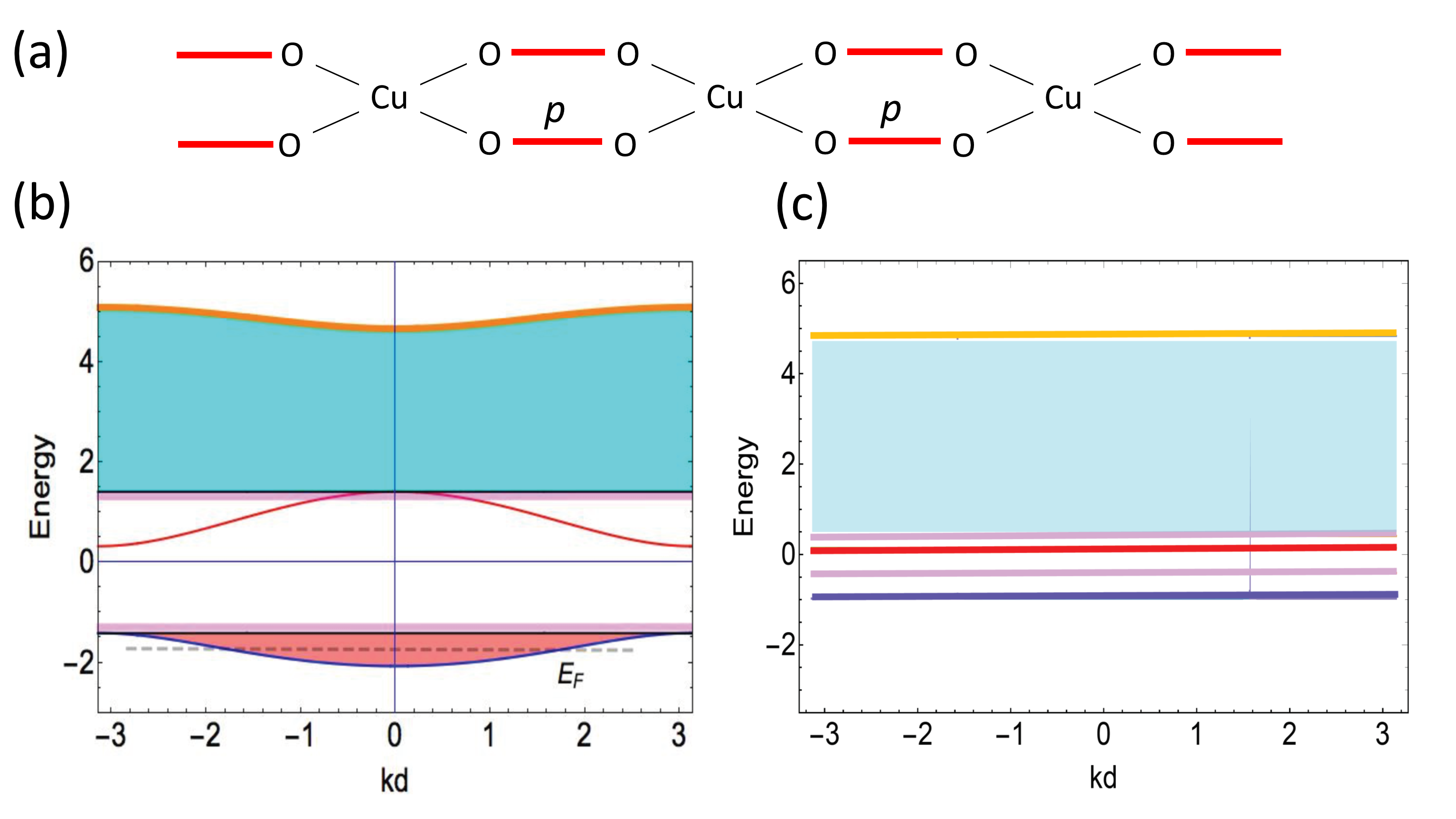}
\caption{\label{bands1} a) A structure of the conducting filament. The unit cell has 5 atoms so there are 5 energy bands;
 b) an energy momentum relation of the conducting filament, for all five existing bands. It is calculated when the value of the polarisation is $p=-0.9 t$.  Note that the lowest energy band has two minima at the boundary of the Brillouin zone (BZ). c) When one polarisation in the unit cell changes sign all 5 bands become completely flat.  Such localisation is related to the formation of Aharonov-Bohm cage\cite{Vidal-1998,Kusmartsev-1995}.
 }
\label{bands}
\end{center}
\end{figure}

The formation of long conducting filaments is proposed to be linked to a condensation of such single spin-orbital polarons. Here many electrons become trapped in a single polarisation well, produced through the polarisation of the oxygen orbitals. The minimum of the Coulomb repulsion between electrons corresponds to a linear potential well, known as polaron string and stripe\cite{Kusmartsev-1999,Kusmartsev-2000,Bianconi-Kusmartsev}. In this way, the electron motion is confined along a linear path across the Cu sites with the same spin orientation and connected by polarised oxygen sites. This type of transport is well described by a form of tight-binding model. An example of the structure of a conducting filament is shown in  Fig. \ref{bands1}a. The electron spectrum obtained under those conditions is presented in Fig. \ref{bands1}b. The unit cell of the filament contains five atoms (one copper and four oxygens) consequently giving rise to five bands.
Notably, there are two flat bands associated with localised polarons (when $E=\pm p$) and another three separate bands related to mobile polarons. The energy-momentum dependence for each of these bands, calculated with parameter value $p=0.9t$, is given in the Fig.\ref{bands}b. The position of the lowest energy band is determined by the value of the hopping integral $t$. Its bandwidth depends on the value $p$ and the Hubbard parameter $U$. When the value $p$ increases from zero, the bandwidth increases. Conversely, when the value $U$ increases the bandwidth decreases. In fact, the combination of parameters $p$ and $U$ determines the size of the gap separating the first band from the second one. The gap vanishes when the value of $p$ reaches its critical value $p= \sqrt{2} t$. The energy of the bottom of the lowest band of the filament (its minimum is at $k=0$)  decreases linearly with induced polarisation $p$. Recently, there have been many reported observations of anharmonicity\cite{Chernikov-2014} or local charge-density wave (CDW) in electron-doped cuprates.\cite{Silva-2016,Campi-2015} These effects can be related to spin-orbital polarons or their ordered/condensed state as described here.

In summary, the new general concepts of conducting filaments and spin-orbital polarons have been introduced here within a theoretical framework. In the proposed scenario, the spin-orbital polarons form excitations of the conducting filamentary state.  
For the first time it was demonstrated that the conducting filaments and the spin-orbital polarons may coexist with an antiferromagnetic phase. The resultant electronic state is manifested as a filamentary spider web of one dimensional correlated electrons superposed on an antiferromagnetic background.
The structure of conducting filaments is reminiscent of an electron liquid crystal or electron nematic\cite{kivelson1998electronic}. The filaments form a microscopic electronic analogy to conventional liquid crystals. Moreover, the presence of spin-orbital polarons gives an additional degree of freedom to this picture. At high temperatures the filamentary web transforms into separate spin-polarons. In other words the electron liquid crystal melts into a viscous electronic, possibly, Luttinger liquid\cite{Kusmartsev-1992,Kusmartsev-1993}. While at low temperatures and small electron doping the filaments and spin-orbital polarons can condense into a spider web that coexists with the AFM order. For electron-doped La$_2$CuO$_4$ this will occur at critical doping level $x<0.08$. The application of external magnetic field orders the filaments and makes the charge carriers more mobile, resulting in a combination of negative and positive magnetoresistance effects that are often observed in electron-doped cuprates\cite{Heshan-2017}. The filamentary structure may also naturally describe the creation of a two-phase system important for the  appearance of the magnetoresistance effects.\cite{bulgadaev2005large}. Therefore, the electron-doped side of the phase diagram of cuprates can be consistently explained using these novel phenomenologies. The key of all these effects described here is polarisation of oxygen orbitals. The analogous important role of oxygen atoms have been also noticed for the  vortex or magnetic flux trapping and paramagnetic Meissner effect 
observed in the hole-doped cuprates\cite{Rykov-1997,Kusmartsev-1992,Rykov-1999}. The similar frustrated electronic phase separation as described in this paper has been recently observed by using scanning micro x-ray diffraction and EXAFS methods in cuprates and iron based superconductors showing the emergence of a non Euclidean  hyperbolic space  for filamentary pathways at percolation \cite{Campi-2015,Campi-2016}. The ideas proposed here may also offer new insights into the origin of superconductivity and other electronic instabilities reported in these systems.

Acknowledgements: The authors are very grateful to E.I.Rashba and B. Fine for illuminating discussions. The support and hospitality of Center for Theoretical Physics of Complex Systems, Daejeon, South Korea, where a part of this work has been done, is highly appreciated.


\end{document}